%% file: ulx3.tex
\documentclass[usenatbib, a4paper, useAMS, referee, lscape]{mn2e}

\usepackage{amsmath}                   
\usepackage{amssymb}                   
\usepackage[UKenglish]{babel}    
\usepackage{rotating}                  
\usepackage{graphicx}                  
\usepackage{multirow}
\usepackage[colorlinks=true]{hyperref} 
\hypersetup{urlcolor=blue}             

\newcommand{\be}{\begin{equation}}
\newcommand{\cm}{{~\rm cm}}

\newcommand{\ee}{\end{equation}}

\newcommand{\ergs}{{{~\rm erg}\; \rm s^{-1}}}
\newcommand{\keV}{{~\rm keV}}
\newcommand{\km}{{~\rm km}}

\newcommand{\msun}{{~M_\odot}} 
\newcommand{\Mpc}{{~\:\rm Mpc}}
\newcommand{\s}{{~\rm s}}
\newcommand{\yr}{{~\rm yr}}

\title{Chandra  observations of the ULX N10 in the Cartwheel galaxy}

\author[Pizzolato Wolter \& Trinchieri]{
	Fabio Pizzolato
	\thanks{E-Mail: fabio.pizzolato@brera.inaf.it}
	\and
	Anna Wolter 
 	\and
	Ginevra Trinchieri
	\\
	INAF--Osservatorio Astronomico di Brera
	Via Brera 28 20121 Milano--Italy
}

\begin{document}

\date{\today}

\pagerange{\pageref{firstpage}--\pageref{lastpage}} \pubyear{2010}

\maketitle

\label{firstpage}

\begin{abstract}

The Cartwheel galaxy harbours more  Ultra--Luminous X--ray  sources 
(ULXs) than
any other galaxy observed so far, and as such it is a particularly interesting target to study them.
In this paper we analyse the three  {\sl Chandra} observations of
the brightest ULX (N10) in the Cartwheel galaxy, 
in light of current theoretical models suggested to explain such
still elusive objects. 
For each model we derive the 
relevant spectral parameters. Based on self--consistency arguments
we can interpret N10 as an accreting binary system
powered by a~$\sim100\msun$ black hole.
A  young supernova strongly interacting with its surroundings is a likely
alternative, that can be discarded  only with the evidence of a flux increase
from future observations.
 
\end{abstract}




\section{Introduction}
\label{s-intro}

The Ultra--Luminous X--ray  sources (ULXs) are a class of  very bright
($L_X\simeq 10^{39}-10^{41}\ergs$),
point--like X--ray sources detected off the nuclei in several
galaxies \citep{Fab06}. Although they were discovered more than 20 years ago  
(by the {\sl Einstein} X--ray satellite:
\citealp{Lon83,Fab89}),
their nature remains unclear.
If the ULXs are powered by accretion, their luminosity exceeds
the Eddington limit  for  a stellar--sized object, sometimes by a 
factor~$\sim1000$. 
This has led some authors \citep{Col99} to postulate that the 
ULXs are black holes with masses 
$10^2-10^4\msun$, intermediate  between the black holes expected from the 
final evolution of massive stars
and the super--massive black holes 
of $10^6-10^9\msun$ powering the Active Galactic Nuclei.
Possible formation scenarios of these intermediate mass black holes 
(IMBH) are discussed in \citet{Mil02} (formation in globular clusters), and
\citet{Por04} (in young super--massive star clusters). 
An alternative view places the ULXs in the more familiar realm of
stellar--sized black holes. According to \citet{Kin01}, 
the ULXs are black holes of $M_\bullet \simeq 10\msun$, accreting
mass from a disc, at a rate above their Eddington limit.
In this regime, the inner accretion disc is thick, and may
collimate  the outgoing radiation. The anisotropy 
leads the observer to overestimate the luminosity
(and the mass of the hole) by a large factor, up  to $10-100$.
Alternatively, some models suggest that
the ULXs may actually  radiate above the Eddington limit.
They are slim disc models \citep{Ebi03}, photon bubble dominated
discs \citep{Beg02, Beg06} and two--phase radiatively efficient discs
\citep{Soc06}. A combination of beaming and super--Eddington accretion is also
possible \citep{Pou07, Kin08, Kin09}.

More recently,  a third possibility has been emerging, namely that the ULXs
are black holes of $30-90\msun$ produced by the final evolution of
massive stars in a metal poor environment \citep{Zam09}.
In such an environment the black hole's progenitor does not
loose much of its original mass by the action of line--driven winds,
and might collapse directly into a black hole without exploding as a supernova,
producing more massive  black holes than in a metal rich environment.

The accretion--powered models do not exhaust the possible scenarios able to explain the nature of the ULXs. 
In a  sample of $154$ ULXs observed with {\sl Chandra},  \citet{Swa04} 
argue that about~$20\%$ of them may be well described  by a
thermal model, consistent with young supernova remnants  strongly
interacting with the surrounding environment.
In this interpretation, the Eddington limit is  clearly
not an  issue, and indeed the ULXs have X--ray luminosities 
comparable to those of the brightest supernovae ($L_X\sim 10^{41}\ergs$,
\citealp{Imm03}).

The Cartwheel galaxy is a peculiar nursery of ULXs.  This galaxy
underwent a collision with a smaller galaxy about $10^8\yr$ ago
\citep{Hig96, Map08}. This episode
not only conferred the galaxy its peculiar shape, but also triggered a
massive star formation in its ring. The Cartwheel harbours the largest number of
ULXs than any other galaxy: 15 sources are more luminous  than 
$10^{39}\ergs$ \citep{Wol04}, and at least some of them  are known to be
variable \citep{Cri09}.
In this paper we  address the properties of the ULX  labelled N10 by \cite{Wol04}.
In  the {\sl Chandra} observation of~2001 it was the brightest  ULX in the Cartwheel
(and also among the brightest known ULXs),
but  a couple of  subsequent  {\sl XMM--Newton} observations taken in 2004 and 
2005 showed that its luminosity is fading \citep{Wol06}.
Two additional {\sl Chandra} observations were performed in~2008 
to follow the variability pattern of this ULX. We present here the
{\sl Chandra} observations taken in
2008 and compare them with that of 2001. 
The {\sl ACIS} angular resolution is instrumental 
to reduce to a minimum the possible contamination of N10
from the surrounding diffuse gas  and the neighbour sources crowding the Cartwheel
ring.
In our analysis we do not use the {\sl XMM--Newton} observations,
since the analysis of the variability with {\sl XMM--Newton} data
was already published by \citet{Wol06}.
In addition, the use of {\sl XMM--Newton}
data   would require specific modelling of the 
contamination of the  
neighbouring sources and the Cartwheel's ring 
due to the relatively large PSF of {\sl XMM--Newton}, incrementing
the uncertainties in the derived spectral parameters.

Our main goal is to compare the spectral parameters of N10
to the theoretical models  put forward to explain the ULXs.
In particular, we shall discuss several accretion models
and the supernova model. Although N10 is a bright source, its
large distance ($D=122\Mpc$, \citealp{Wol04}
and references therein) limits the number of collected photons, 
preventing us  from rejecting or confirming a spectral 
model on purely statistical grounds.

For this reason, we shall assess the likelihood of any model
more in the light of its own self--consistency than of its
statistical evidence.


The outline of the paper is the following.
In section~\ref{s-prep} we  describe the {\sl Chandra} 
data and  their preparation. In Section~\ref{s-plaw}
we present a simple spectral model of N10, which will be a thread 
for a more detailed  analysis. The accretion and supernova models are
presented in section~\ref{s-accretion} and~\ref{s-supernova},
respectively. 
Finally, we discuss and summarise our results in section~\ref{s-sum}.

\section{Data Preparation}
\label{s-prep}

{\sl Chandra} observed the Cartwheel once in 2001 and
twice in  2008. All the observations were carried out with the 
back--illiminated (BI) chip  {\sl ACIS-S3}. 
This  detector was  operated in {\sl FAINT} mode in the first
observation (no.~2019) and in {\sl VFAINT} mode in the remaining observations 
(nos.~9531 and~9807).

New level=2 event files have been re-created with  {\sl CIAO-4.1.2} 
in order to work with a homogeneous data set.
The script {\sl wavdetect} was used to detect the sources within the CCD field
of view. No observation was seriously contaminated by flares; to check this
we have first extracted a light curve from the source--free areas
in the spectral band $0.3-10\keV$. The 
{\sl Chandra} script {\sl lc\_sigma\_clip} used this curve to excise the 
periods with a count rate more than $3$~sigmas  above the average. 
This correction excludes but a small fraction of the observations, as shown in 
the summary Table~\ref{t-data}.

The spectra of N10  were  extracted from a circle of
radius $R=2.7''$, and the background spectrum 
from a source--free region near the source.
The chosen extraction circle includes virtually all the photons from the source.
In all our analysis we fit the spectra in the window $0.3-7\keV$, where
the signal to noise ratio is higher. 
In this window the first spectrum
(observation no.~2019) contains $400$~photons,
the second one (observation no.~9531)
$173$ photons  and the third 
(observation no.~9807) only $59$ photons before  background subtraction. 
There are two contributions to the background: 
instrumental and  due to the diffuse gas of the galaxy ring
and to unresolved low luminosity sources. Both of them give a negligible
contamination  on account of the small extraction 
area: we estimate
that the instrumental background  
contributes for less than~$2$ photon in  each observation, and
$3-5$ photons come  from the gas of Cartwheel galaxy's ring.

On account of the relatively poor statistics of our data, 
we would not like to loose spectral resolution by  binning the data
to the standard minimum value of  $\sim 25$~counts/bin required by  a consistent
use of  the $\chi^2$ statistics  (see e.g. \citealp{Cas79}).
For this reason we choose to bin the spectra  with a minimum of $5$~counts/bin
and analyse them  with the implementation of the Cash  statistics 
(called {\em cstat}) provided in the version~12.5.1 of {\sc XSPEC}.  
The Cash statistics 
assumes that the counts are distributed according to a Poisson law,
which does not allow to subtract the background  from the spectra
\citep{Cas79}.   In our case the background contribution is small,
and we simply neglect it.

The  {\em cstat} statistics
differs from the canonical Cash statistics because the value of {\em cstat}
provides a goodness of fit (similar to the $\chi^2$ statistics), if
each spectral bin contains at least $5$ counts. The fit is considered
acceptable if the value of the ``reduced'' {\em cstat} (i.e. the ratio of
{\em ctstat} to the number of degrees of freedom of the model)
is close to one (\citealp{Arn96}
\footnote{See also the {\sc XSPEC} manual page
http://heasarc.gsfc.nasa.gov/docs/xanadu/xspec/manual/XSappendixCash.html.}).
We also check that the distribution of residuals 
 in any spectral model is not skewed and does 
not show peculiar features.

All the statistical uncertainties of the best--fitting parameters
are quoted at $90\%$ confidence level.
The X--ray luminosities of the source 
are  consistently calculated in the spectral band $0.5-10\keV$, and they are
corrected for absorption.

\section{The Power Law Model}
\label{s-plaw}

\begin{figure}
\begin{center}
\includegraphics[width=110mm,angle=-90]{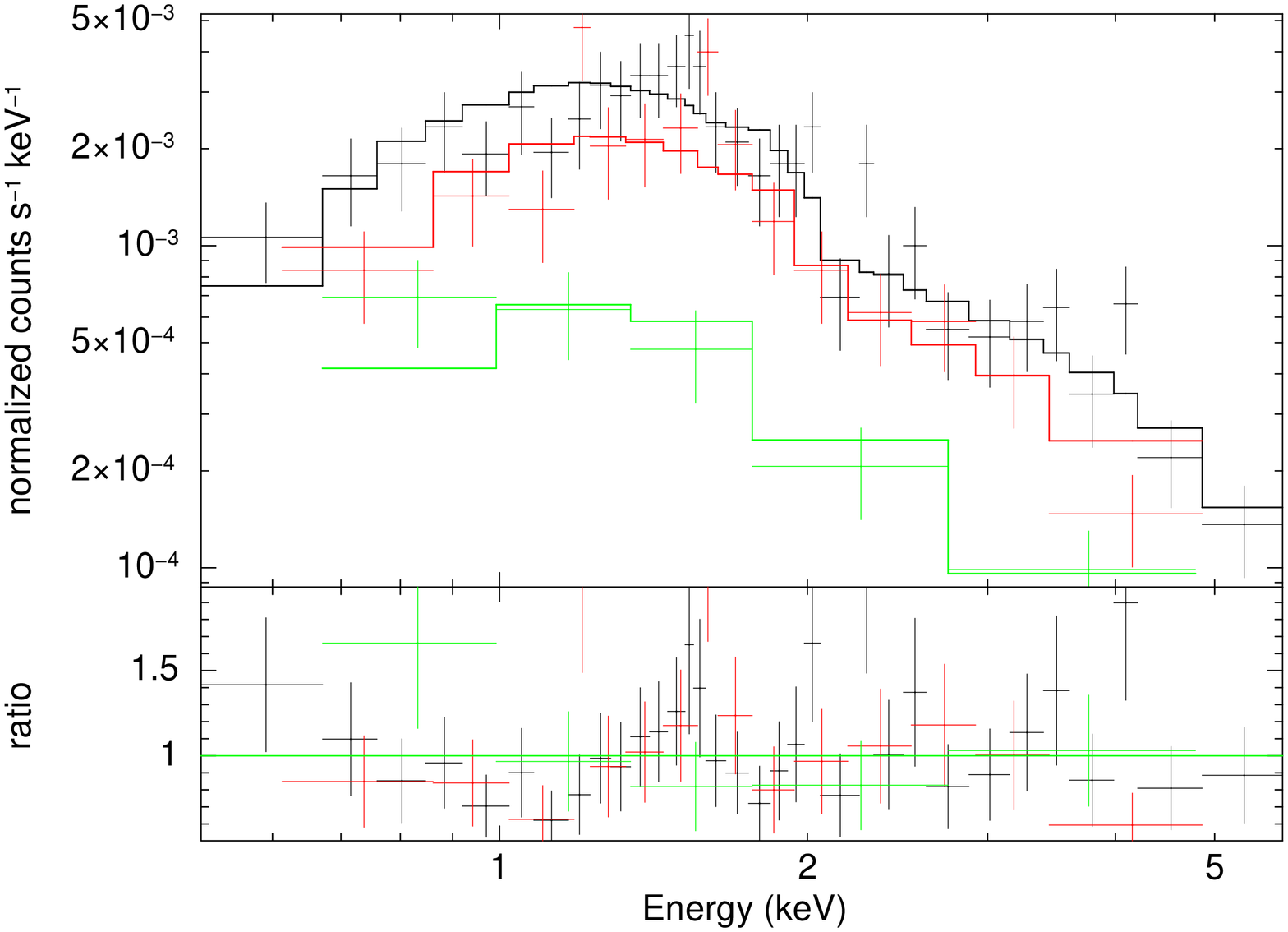}
\end{center}
\vskip12mm
\caption{\label{f-pl}
Data and residuals for the best-fit absorbed power law for the 
three {\sl Chandra} observations. Top (black)  
obs. no.~2019, middle (red) no.~9531,  bottom
(green)  no.~9807.
}
\end{figure}

The count rate between the three observations varies.
In order to avoid complications in comparing count rates of 
observations  taken in {\sl FAINT} and {\sl VFAINT}
modes, we evaluate this variability 
with a simple absorbed power law model (${\tt wabs*powerlaw}$). 
Despite its
simplicity,  it is informative of some basic parameters useful to 
our further analysis. Separate fits of each observation return 
comparable values of the absorption column $n_H$ 
and the spectral photon index  $\Gamma$.
For this reason  we tie $n_H$ and $\Gamma$ of each data set 
to a common value and fit them simultaneously. 
Table~\ref{t-plaw} summarises the results of the fit, and 
Figure~\ref{f-pl} shows the model and its residuals.
The  value of the C-statistics is $C/{\rm dof}=100.8/101$.
The best fitting absorption column is
$n_H=3.7_{-1.0}^{+1.0}\times 10^{21}\cm^{-2}$, and 
the spectral photon index is $\Gamma=1.88_{-0.24}^{+0.25}$, both 
consistent with those  presented by \citet{Wol04}.
The value of $n_H$ is higher than Galactic, but consistent both
with other X-ray measures in the Cartwheel and with the intrinsic 
absorption in the optical band
(see \citealp{Wol04}).

The intrinsic luminosities in the spectral band $0.5-10\keV$  is 
$1.2\times 10^{41}\ergs$, 
$8.6\times 10^{40}\ergs$ and 
$2.8\times 10^{40}\ergs$
for the first, second and third observation respectively. 
Compared to the Eddington luminosity  of a $1\msun$ collapsed object,
these  luminosities seem to imply a black hole of $\gtrsim 200\msun$.
It is perhaps worth to remark that 
the Eddington luminosity is {\em bolometric}, while 
in the present paper we refer to the luminosity in the X--ray band.
This is an  approximation of the 
bolometric luminosity  within a 
factor $\gtrsim 2$.

Formally, a power law model provides an adequate description of the
{\sl Chandra} data sets, and more refined models are not statistically required.
Nevertheless, we would like to know whether these observations
may somehow  constrain  some of 
the theoretical models suggested for the ULXs. The following sections 
aim at this goal.

\section{The Accretion Models}
\label{s-accretion}

In the framework of an accretion scenario, the engine of N10 is a
black hole  accreting  mass from a donor star through a disc. 
In this section we  present the results of the spectral analysis of 
three accretion models:
i)~a multi--colour disc ({\tt diskbb} in the language of {\sc XSPEC}),
ii)~a model of a disc around a maximally rotating Kerr black hole 
($\tt kerrd$ in {\sc XSPEC}) 
and finally iii)~a \citet{Kaw03} 
slim disc model
\footnote{
Slim disc models are not available in the main release of the package, but only
as tabular models to be downloaded separately
from the URL 
\url{http://heasarc.gsfc.nasa.gov/docs/xanadu/xspec/models/slimdisk.html}.}.
All models are acceptable on statistical grounds and they all 
return similar values of the C-stat.
For an easier comparison, we have summarised the results of these models 
in Table \ref{t-acc}.
We also shall discuss the hyperaccretion model
suggested by \citet{Kin01}, but since no spectral model is available for
this, our discussion will be mainly qualitative.

\subsection{The multicolour disc}
\label{s-multicolour}

The first model we consider is a multicolour disc (MCD) modified 
by the 
interstellar absorption.
The two free  parameters of the model are the effective temperature of the inner rim
$T_{\rm in}$ and the apparent inner disc radius $R_{\rm in}$, 
related to the actual inner radius $R_{\rm disc}$ through the relation
\citep{Kub98}
\be
\label{e-corr}
R_{\rm disc} = \xi \; f^2 \; R_{\rm in},
\ee
where $\xi\simeq 0.412$ is a correction factor that takes into account that the
maximum temperature $T_{\rm in} $ does not occur exactly at the inner edge
of the disc \citep{Kub98}. 
For typical parameters,  the  disc optical opacity  is dominated by electron scattering. The electron scattering may also  Comptonise  the emerging spectrum, and   
this effect is taken into account by introducing the so called  
hardening factor $f$, i.e. the ratio  $f\equiv T_{\rm col} /T_{\rm eff}$ 
between the colour temperature $T_{\rm col}$ 
and the effective temperature $T_{\rm eff}$.
The value of $f$ is insensitive to the disc viscosity 
and the mass of the accretor, but is sensitive to the mass accretion rate.
For a mass accretion rate close to  the Eddington limit
the value of $f$ lies  in the range
$f\simeq 1.7-2.0$ \citep{Shi95}.
\citet{Kaw03} argues that if the accretion rate largely exceeds 
the Eddington limit (${\dot M}\, c^2/L_{\rm Edd}> 100$),
the hardening factor is somewhat larger, in the range
$f\simeq 2.3-6.5$.
Introducing the  radiation efficiency 
\be
\label{e-eff}
\eta=L/{\dot M} c^2,
\ee  
this limit translates into 
\be
\label{e-kaw}
L>100\, \eta\: L_{\rm Edd}.
\ee
For a disc accreting on a Schwarzschild black hole $\eta\simeq 0.06$
\citep{Fra02}, so \citeauthor{Kaw03}'s hardening parameter 
applies  if $L\gtrsim 6\, L_{\rm Edd}$.

The normalisation of the model only involves  geometrical parameters,
so we link the normalisation of the three spectra to a common value, and
leave the  disc temperatures free to vary. Since the absorption columns
inferred from each spectrum are consistent, we also link them to a common value.
The  interstellar absorption column is 
$n_H=1.8_{-0.6}^{+0.6}\times 10^{21}\cm^{-2}$.
The disc inner temperature is 
$T_{\rm in}=1.33_{-0.17}^{+0.22}\keV$ for the first observation;
$T_{\rm in}=1.21_{-0.16}^{+0.17}\keV$ for the second  observation and
$T_{\rm in}=0.89_{-0.11}^{+0.14}\keV$ for the third  observation.
The normalisation of the model is $K = 7.4_{-3.3}^{+5.2}\times 10^{-4}$, and  
is related to the apparent inner radius $R_{\rm in}$ through the expression
\be
R_{\rm in} = 1.22\times 10^2 \km\; \left(\frac{K}{10^{-4}}\right)^{1/2} \; \left(\frac{D}{122\Mpc}\right) \; \bigl(\cos\theta\bigr)^{-1/2},
\ee
where $\theta$ is the inclination of the disc with respect to
the line of sight  
($\theta=0$ is face--on), and $D$ is the distance to the source.
Inserting the appropriate value of $K$ we find
\be
R_{\rm disc} \simeq 3.3_{-0.8}^{+1.0} \times 10^2 \; \xi\; f^2 / (\cos\theta)^{1/2}\; \km.
\ee
If the inner edge of the disc is located at the last stable orbit of a
Schwarzschild black hole, this implies
\be
\label{e-mbh}
M_\bullet \simeq 61_{-12}^{+18} \msun\; \left(\frac{f}{2.0}\right)^2 \;
\left(\frac{\xi}{0.41}\right) \; \frac{1}{(\cos\theta)^{1/2}}.
\ee
The resulting disc luminosities are
$L_1=4.1\times 10^{40}\ergs$, 
$L_2=2.7\times 10^{40}\ergs$ and
$L_3=8.0\times 10^{39}\ergs$, for the first, second and third observation,
respectively.
All these values exceed  the   Eddington luminosity 
$L_{\rm Edd}\simeq 7.9 \times 10^{39}\ergs$ 
for a black hole of~$M_\bullet\simeq 60\msun$ by a factor
$L/L_{\rm Edd} \lesssim 5$,  consistent with  our
choice of the hardening parameter $f=2$.

A long--standing problem of the application of MCD models
to the ULXs is that they tend to  overestimate the disc's inner temperature 
$T_{\rm in}$. Indeed, the luminosity  of a MCD is \citep{Kub98}
\be
\label{e-lbol}
L_{\rm bol} = 4\,\pi\: \left(\frac{R_{\rm disc}}{\xi}\right)^2 \:
\sigma_{\rm SB} \; \left(\frac{T_{\rm in}}{f}\right)^4
\ee
\citep{Mak00}.
If the temperature is too high 
(for the same luminosity and the other parameters), the disc radius
$R_{\rm disc}$ is underestimated, so is the black hole's mass.
A possible explanation for this effect is the presence of a hot
corona embedding the disc \citep{Sto06}: a single MCD model applied to
such system 
would return spuriously high disc temperatures.
We investigate how the presence of a hot corona would affect our 
estimate~\eqref{e-mbh} of the black hole mass.
We describe  N10 with a Comptonised disc emission,
where   the temperature of the Compton seed photons is set equal to 
the disc inner temperature.
On account of our rather low statistics,
the temperature of the hot corona is not well constrained, and
it  has been fixed to $k_B T_C=50\keV$, while
its  optical depth  has been left free to vary. 
The best-fitting value 
of the Compton thickness is $\tau\simeq 0.5_{-0.2}^{+0.4}$:
this results strongly depends on the
temperature $T_C$, with cooler coronae returning   larger optical depths.
The temperature of the disc, however, is not  sensitive to the 
precise values of $T_C$ and $\tau$.
As noted by \citet{Sto06}, a Compton--thick ($\tau>1$) corona
obscures the inner region of the disc, and in this case the MCD
parameters cannot provide  reliable  estimates of the black hole's mass
(see also \citealp{Gla09}).
The disc temperature of the three observations are respectively
$T_{\rm in}=0.53_{-0.12}^{+0.11}\keV$,
$T_{\rm in}=0.36_{-0.09}^{+0.16}  \keV$ and 
$T_{\rm in}=0.09_{-0.04}^{+0.05} \keV$. 
The normalisation of the disc component is 
$K= 7.6_{-3.9}^{+10.6}\; \times 10^{-3}$
(in {\sc XSPEC} units), yielding
\be
\label{e-mbh3}
M_\bullet = 141_{-42}^{+78}\msun \;
\left(\frac{\xi}{0.41}\right) 
\;
\left(\frac{f}{1.7}\right)^2
\ee
Using the above disc inner temperatures we 
infer the luminosity of the disc component  to be
$L_1=9.0\times 10^{39}\ergs$ (first observation),
$L_2=1.5\times 10^{39}\ergs$ (second observation) and
$L_3=5\times 10^{35}\ergs$ (last observation).
The ratio between $L_1$ and the Eddington luminosity  of a black hole
of $M_\bullet\simeq 141\msun$ is~$\sim0.5$, which justifies our 
choice  $f=1.7$ of the hardening parameter.
As expected (see \citealp{Sto06}), a hot corona lowers our estimate of 
disc's temperature and increases 
the black hole's mass with respect to a simple MCD model.

\subsection{The Kerr disc} 
\label{s-kerr}

The second model we consider is the 
spectrum emitted by a disc orbiting a  maximally rotating Kerr black hole.
We fix the distance of the Cartwheel to its  known value, and 
the hardening parameter to~$f=1.7$: the quality of our fit is independent of
this choice.
The inclination of the 
disc with respect to the line of sight has been fixed to $\theta=0^\circ$;
the value of $\theta$ affects the determination of $M_\bullet$,
as we shall discuss at the end of this section. 
We also set the radius of the accretion disc to the last stable orbit  
of a  maximally rotating Kerr BH ($\simeq 1.235\, G \,M_\bullet/c^2$), and  
the disc outer radius to its default value 
(the model is insensitive to this parameter).
The only free parameters of the model (apart from the hydrogen
column density) are  the mass $M_\bullet$ of the central object
and its  mass accretion rate $\dot{M}$.
In the fitting procedure we link all the
parameters for the three observations except  the mass accretion rates.

The fit  returns the 
absorption column $n_H=2.2_{-0.6}^{+0.6}\times 10^{21}\cm^{-2}$,
and the mass of the central object results
\be
\label{e-mass}
M_\bullet = 92.8_{-27.7}^{+32.4}\msun.
\ee
The luminosity of the source  is 
$L_1=4.3\times 10^{40}$ in the first observation,
$L_2=2.9\times 10^{40}$ in the second, and 
$L_3=8.5\times 10^{39}$ in the last.
The Eddington luminosity associated to the mass~\eqref{e-mass} is
$L_{\rm Edd}=1.2 \times 10^{40}\ergs$. These values must be checked for their
consistency with the adopted hardening parameter. A Kerr maximally rotating 
black hole has an accretion efficiency   $\eta\sim 0.4$
(\citealp{Fra02}, see Equation~\ref{e-eff}). \citeauthor{Kaw03}'s larger 
values of $f$ then apply if $L\gtrsim 40\, L_{\rm Edd}$. 
In all our observations,
the Kerr disc model returns
$L/L_{\rm Edd}\lesssim 3.5$, so  the adopted hardening
parameter is consistent with the inferred $M_\bullet$ and the 
relatively low mass accretion regime.

\begin{figure}
\begin{center}
\includegraphics[width=110mm,angle=-90]{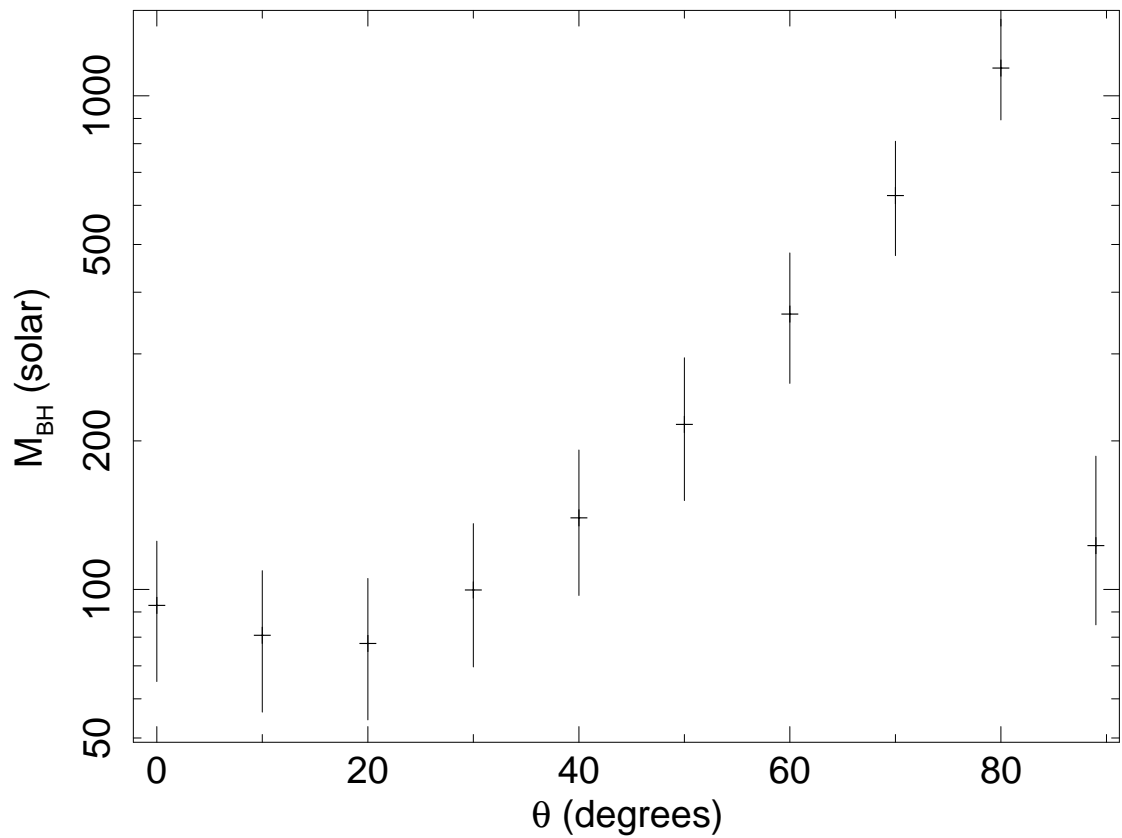}
\end{center}
\caption{\label{f-i_m}
The best--fitting mass $M_\bullet$ of the model ${\tt wabs*kerrd}$ 
as a function of the inclination $\theta$ between the disc and the line of sight.
The error bars represent the $90\%$ uncertainties calculated 
with the free parameters.
}
\end{figure}

Before ending this section we need to  address 
the effect of the  disc inclination $\theta$ 
on the estimate of the mass $M_\bullet$.
The value of $\theta$ does not affect the quality of the fit, but the 
mass is sensitive to it, as shown in Figure~\ref{f-i_m}
(see also \citealp{Hui08}).
Low inclination angles imply lower $M_\bullet$, which 
(for $f=1.7$) attains a minimum value of~$\sim 77\msun$ for $\theta=20^\circ$.  
If the disc is seen almost edge--on, on the other hand, 
$M_\bullet$ grows up to $M_\bullet\simeq 10^3\msun$. 
This trend of $M_{\bullet}$ with the viewing angle is also 
found  in the multicolour discs considered in the
previous section (in particular, see Equation~\ref{e-mbh}).

To summarise, the Kerr disc model is consistent with a black hole of
$M_\bullet\simeq 90\msun$, similar (within the uncertainties)
to what  inferred from the  Comptonised multicolour disc model.
Higher masses are not excluded, if the disc is  highly tilted
with respect to the line of sight. Also
a higher hardening factor would increase $M_\bullet$, but higher
values of $f$ are  not required  by
the inferred accretion regime.

\subsection{Slim Discs}
\label{s-slim}

The  multicolour disc model
assumes that the disc radiates locally as a black body.
If the accretion rate largely exceeds the Eddington limit,
this hypothesis is not appropriate.
In such an accretion regime  the heat content of the disc  is trapped
and dragged to the black hole's horizon  before it is radiated.  
For this reason, {\em slim discs}
are radiatively inefficient, and their properties significantly differ
from those of the standard thin discs (see e.g. \citealp{Abr88, Fra02}).
In this section we investigate whether the data support
the possibility that N10 is powered by  a slim disc 
accreting above its Eddington limit.

\begin{figure}
\begin{center}
\includegraphics[width=110mm,angle=-90]{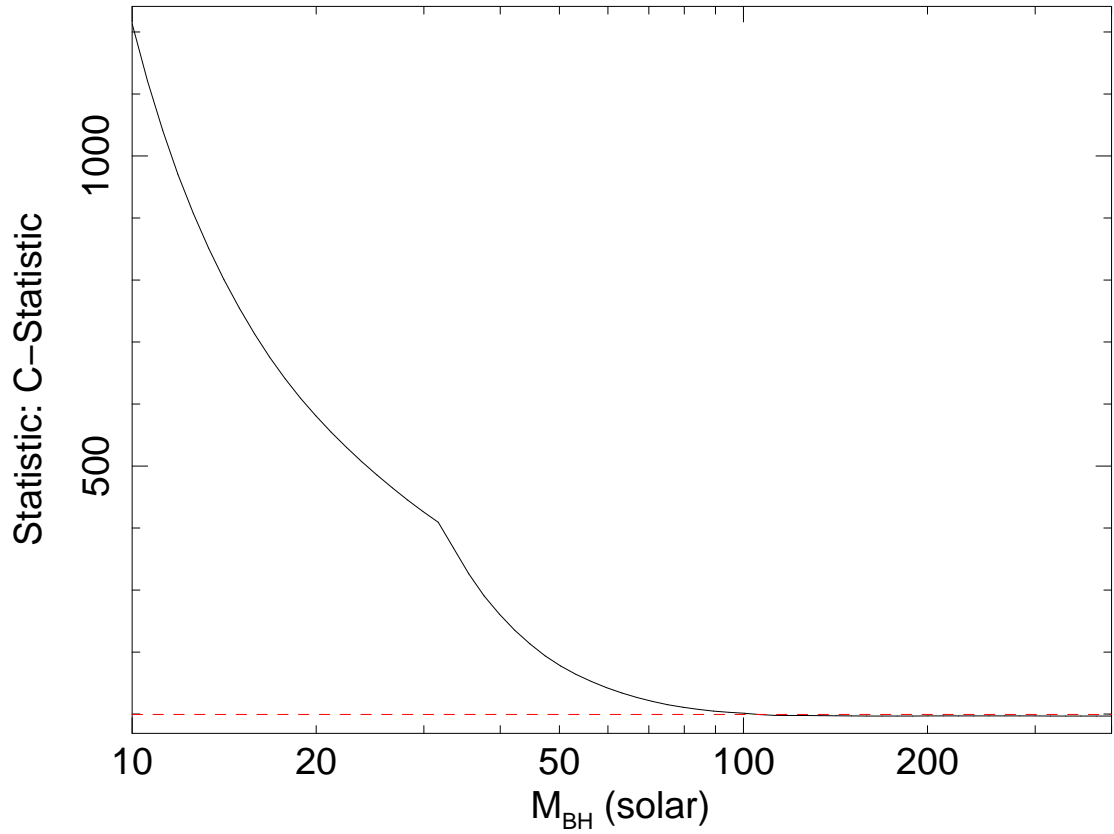}
\end{center}
\caption{\label{f-kawmass}
The dependence of the fit C-statistics on the black hole mass
in the slim disc model by \citet{Kaw03}. Since
the C-statistics curve is flat for $M_\bullet\gtrsim 80\msun$,
the best--fitting mass of the hole can be anywhere above this value.
}
\end{figure}

In {\sc XSPEC} two slim disc models  are available:
we adopted the model based on the calculation by \citet{Kaw03},
with 
the  mass of the central object,
the  accretion rate and the viscosity~$\alpha$ as free parameters. 
The spectrum is calculated taking into account the 
local Comptonisation and the relativistic effects.
\citeauthor{Kaw03}'s model has been applied to fit the {\sl XMM--Newton} spectra of
some ULXs (see e.g. \citealp{Oka06, Vie06}), returning black hole masses
of few tens of~$\msun$ accreting at super--Eddington regimes.
In our fit we have linked the interstellar absorption column,
the hole mass and the $\alpha$~parameter between the observations. 
The mass accretion rate is
independent for each observation. 
The fitting procedure is unable to constrain the black hole's mass. 
The confidence regions are open for
$M_\bullet\gtrsim 80\msun$ (Figure~\ref{f-kawmass}), and the  model is
consistent with any value of  $M_\bullet$  above this limit. 
The absorption column is
$n_H=3.6_{-0.9}^{+0.7}\times 10^{21}\cm^{-2}$,
while the viscosity parameter and the BH mass are 
$\alpha=0.56_{-0.37}$ and 
$M_\bullet= 495_{-340}\msun$ respectively.
The mass accretion rates  for the three observations are 
${\dot M}_1=39.4_{-23}$, 
${\dot M}_2=21.4_{-12.0}^{+523}$,
 ${\dot M}_3=5.1_{-1.2}^{+79}$,
all in units of ${\dot M}_{\rm Edd}\equiv L_{\rm Edd}/c^2$.
Assuming the conversion factor 
${\dot M}_{\rm Edd}=1.3\times 10^{19}~{\rm g}\s^{-1} (M_\bullet/100\msun)$
used by \citet{Kaw03}, the best fitting accretion rates are 
$\dot{M}_1=2.5\times 10^{21}~{\rm g}\s^{-1}$,
$\dot{M}_2=1.4\times 10^{21}~{\rm g}\s^{-1}$,
$\dot{M}_3=3.3\times 10^{20}~{\rm g}\s^{-1}$.
It is not possible to set upper limits to the 
the parameters $\alpha$, $M_\bullet$
and $\dot{M}_1$, since the error calculation
hits  the limits of their tabulated values 
($\alpha=1$, $M_\bullet=1000\msun$
and $\dot{M}=1000~{\dot M}_{\rm Edd}$, respectively).

One  {\em caveat} is in order about the consistency of
the  application of  slim disc models  to  N10.
The radial temperature profile of slim discs  varies
as $T\propto R^{-1/2}$ instead of $T\propto R^{-3/4}$
characteristic of  thin multicolour discs \citep{Wat00}.
The radial dependence of $T_{\rm in}$ in N10 may be checked with 
a multicolour disc ({\tt diskpbb} in {\sc XSPEC})  where
the temperature index $p$  (defined by $T(R)\propto R^{-p}$)
is a free parameter.
The index $p\simeq 0.5$ has been found in some ULX spectra successfully
fitted with a slim disc model \citep{Oka06, Vie06}.
In our case the model {\tt wabs*diskpbb} is comparable to 
the others and since its  
best--fitting parameters are very 
similar to those derived  from a standard multicolour disc, we do not present
them.
Our  best fitting value is $p\simeq 0.74_{-0.19}^{+3.66}$ 
close to the standard value~$0.75$, but 
slightly inconsistent with the  slim disc  value $p=0.5$.

\subsection{The ``Hyperaccretion" model}
\label{s-hyper}

The last accretion model we consider for N10 is the so--called 
``hyperaccretion" scenario \citep{Kin01, Kin08, Kin09}.
This model explains the high luminosity of the ULXs with 
a combination of a high accretion rate (close to the Eddington limit) 
on a stellar mass black hole, and  
a mechanical beaming due to the accretion stream itself. 
 
The apparent X--ray luminosity (Equation~(10) in \citealp{Kin09}) of the source 
results 
\be
\label{e-hya}
L\simeq 2.2 \times 10^{36} \ergs \; \left(\frac{M_\bullet}{\msun}\right) \; 
\left(\frac{\dot{M}}{\dot{M}_{\rm Edd}}\right)^2
\; 
\left[1 + \ln\left(\frac{\dot{M}}{\dot{M}_{\rm Edd}}\right)\right] 
\ee
where 
$\dot{M}$ is the mass transfer rate from the donor star and
$
\dot{M}_{\rm Edd} = L_{\rm Edd} / \eta \, c^2 
$
is the Eddington--limited mass transfer, dependent on the efficiency 
$\eta$ of the X--ray production  in the accretion process, defined 
by Equation~\eqref{e-eff}. 

One of the most relevant features of this model is that the  beaming 
factor scales as ${\dot M}^{-2}$. This causes the 
bolometric luminosity to scale with the disc's temperature as 
\be
\label{e-kingcorr}
L\propto T_{\rm in}^{-4},
\ee
opposite to the standard correlation   $L\propto T_{\rm in}^4$.

The prediction~\eqref{e-kingcorr} has  been found consistent with the
$L-T_{\rm in}$ relation  of the soft excess observed in some ULXs
\citep{Fen07, Kaj09, Gla09}. 
We assume that the count rate of N10 correlates with its intrinsic
luminosity in all our observations. 
In principle, it is certainly true that a strong spectral change may alter
(even reverse) 
this correlation, but such a variation of the spectrum of N10
is not supported by our data.
Under this working hypothesis, we measure a positive correlation between the 
inner temperature and  the luminosity, 
at odds with the prediction of the hyperaccretion model.
The hyperaccretion scenario  probably is not the correct explanation for the nature
of N10.

\section{The Supernova Model}
\label{s-supernova}

As discussed in the introduction, accretion power  is not
the only viable explanation for the ULXs' engine.
Alternative models are possible, and young SNe interacting
with their surroundings may explain an appreciable fraction of the
known ULXs \citep{Swa04}.  In this section we
investigate  the SN model for N10, aiming at deriving the spectral
parameters.

The X--ray emission from supernovae may start up to about  one year
after the explosion,
and it is powered by the interaction of the ejecta
with the circumstellar medium (CSM, see e.g. \citealp{Imm03}). 
Shock fronts form as the ejecta expand through the CSM. The leading 
(or ``forward") shock heats the CSM up to $10^9-10^{10}~\rm K$, and gradually an
inner (or ``reverse") shock wave starts  to propagate. 
The density behind the reverse
shock is $5-10$ times higher  than that behind the forward shock, and the temperature
is lower, around $10^7-10^8~\rm K$ \citep{Che94}. 
For this reason, the soft X--ray emission
(below $\lesssim 5\keV$) is
dominated by the reverse shock, and is well modelled by thermal models, while the 
hard emission (say, above $10\keV$) is dominated by the forward shock.

The luminosities  of X--ray supernovae  lie in the range $10^{37}-10^{41}\ergs$,
with the so--called SN~IIn supernovae
at the bright end of the interval.
These are core--collapse supernovae, owing the ``n" in their name to the
presence of several narrow emission lines in their  optical spectrum
\citep{Tur03}. 
These lines are  most likely due to the 
interaction of the ejecta  with a  dense, slow wind emitted by the 
SN progenitor. 
Other supernovae explode in thinner environments, resulting in
dimmer  X--ray luminosties. For this reason, if N10 is actually a young SN,
it is most likely a SN~IIn.

We fit the spectra of N10 with the thermal models  {\tt apec} and {\tt nei},
corrected for interstellar absorption.
The {\tt apec} model assumes that the emitting plasma is in full
collisional equilibrium: there is a dynamic balance between the 
collisional ionisation and the electronic recombination. The {\tt nei} model
does not assume collisional equilibrium, allowing a mismatch between
the collisional ionisation and the (longer) recombination process.
This mismatch may prevail in shocked plasmas, where the electrons were
energised but did not yet settle to collisional equilibrium with the ions
(see e.g. \citealp{Dop03}).
The parameters of the  {\tt nei}  model are very similar to those
derived for the {\tt apec} model (the fit is quite insensitive to
the ionisation age), so we present only the {\tt apec} results.

In our model ({\tt wabs*apec}) the absorption column and the metal abundances 
of the three observations   are linked together. 
Since there is no compelling statistical evidence of a variation of
the temperature, we also link together the temperatures of the three observations.
The results of this model are presented in Table~\ref{t-apec}.
The statistics of the fit is
$C/{\rm dof}=95.9/100$, and the values of the best--fitting parameters are
$n_H=2.8_{-0.6}^{+0.8}\times 10^{21} \cm^2$,
$k_B T=5.1_{-1.6}^{+3.1} \keV$, 
$N_1 = 4.3_{-1.2}^{+0.9} \times 10^{-5}$  (normalisation of the first observation),
$N_2 = 3.1_{-0.9}^{+0.7} \times 10^{-5}$  (normalisation of the second observation), and
$N_3 = 1.0_{-0.2}^{+0.3} \times 10^{-5}$  (normalisation of the third observation). 
The units of these quantities are (for sources at  low redshift)
$10^{-14}\;  \int dV n_e\; n_H / 4\,\pi\, D^2$,
where $D$ is the distance to the source, $n_e$ and $n_H$ are the electron
and proton densities of the emitting plasma, and the integral (known
as {\em emission integral})  is extended
over the volume occupied by the source.
The emission integrals are consistent with the 
size (few~$10^{15}\cm$) and the density ($\sim10^{-15}-10^{-16}~\rm g\cm^{-3}$) expected from the reverse shock of a young supernova remnant few years after the explosion
(see e.g. \citealp{Che03}).

\begin{figure}
\begin{center}
\includegraphics[width=110mm,angle=-90]{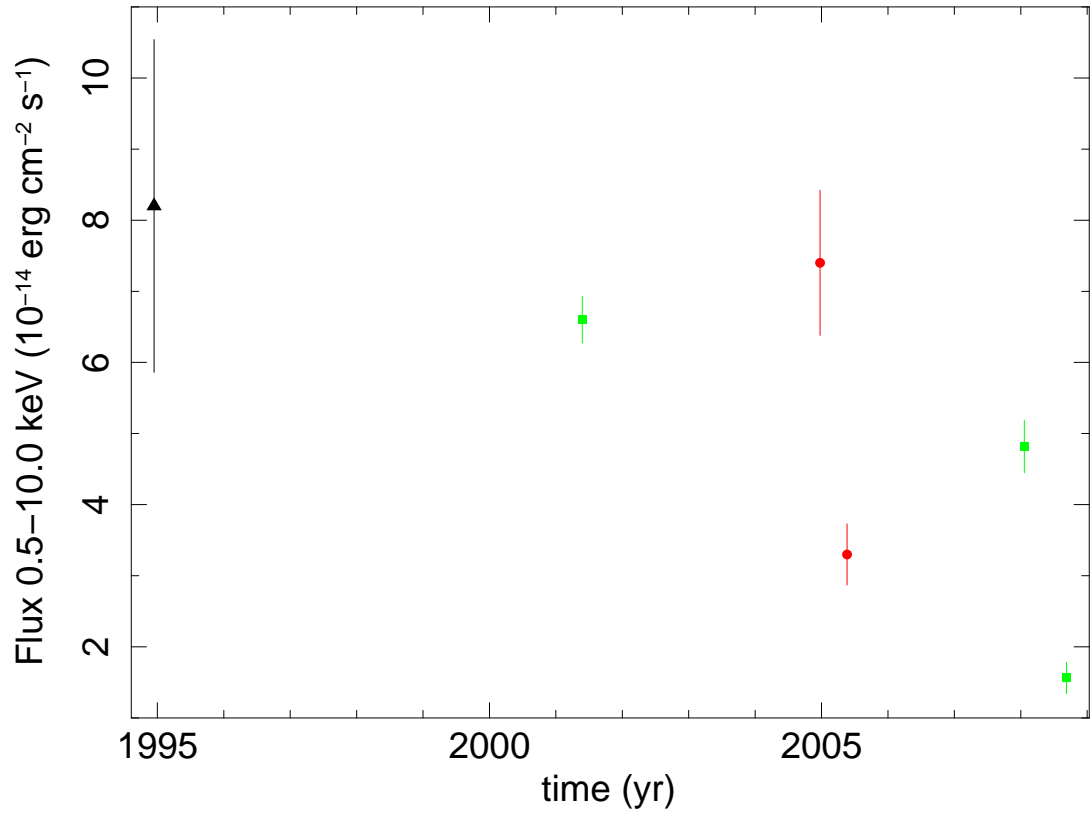}
\end{center}
\caption{\label{f-snlc}
The light curve of the source N10. Fluxes have been calculated with a 
power-law model corrected for the interstellar
absorption. 
The symbols refer to  {\sl ROSAT} data (triangle)
{\sl XMM--Newton} (circles)
and {\sl Chandra} (squares). {\sl ROSAT} and {\sl XMM--Newton} data
are taken from \citet{Wol06}.
}
\end{figure}

The unabsorbed light curve of N10, calculated according to the power law model,
is plotted in Figure~\ref{f-snlc} (see also \citealp{Wol06}). 
The evolution of the observed flux
cannot be fitted by a power law, as observed  in other X--ray supernovae (1986J, \citealp{Hou98}; 1988Z, \citealp{Sch06}; 1998S, \citealp{Poo02}). 
This is not an argument against the SN nature of N10, though, since a simple
power--law decline of the X--ray luminosity has not been reported in other
confirmed SN~IIn. SN~1995N, for instance,
shows a complex light curve,
possibly resulting from the interaction of the ejecta with a complex
circumstellar medium \citep{Zam05}.
The  sudden decrease of the  X--ray luminosity of N10 in 2008 might show 
that the ejecta have passed the dense region of the
pre--supernova wind, and have finally reached a low density  outer environment. 

SN IIn are usually
found in high metal environments (1986J, \citealp{Hou98}; 
1998S,  \citealp{Poo02}).
Indeed, high metal abundances favour the blowing of the line--driven 
strong pre--supernova
winds required to set up the dense circumstellar environment required to 
power a stronger
X--ray emission. The metallicity of the Cartwheel is low as measured from 
the HII regions \citep{Fos77};  the X-ray measure is 
poorly constrained, and consistent with zero.
Therefore the SN progenitor had to be quite massive to blow dense enough 
winds without the aid of a high metallicity.

\section{Discussion  and Summary}
\label{s-sum}

In this paper we considered the spectral modelling of the ULX
N10, located in the Cartwheel galaxy in order to assess
the nature of this source.  
Although the  source is intrinsically very bright,  the spectra have 
few counts on account of the large distance ($122\Mpc$).
For this reason, we are unable to reject (or confirm) a model  on
purely statistical grounds. We first  discussed several accretion models
(multicolour disc around a Schwarzschild or Kerr black hole, slim disc,
hyperaccretion).
 All models indicate 
a black hole of~$\sim 100\msun$, at the high end of the mass distribution
of a black hole 
generated by stellar evolution, possibly in a metal poor environment.
This result is consistent with  the 
conclusions of~\citet{Map09} and~\citet{Zam09} that stellar evolution in a  metal depleted 
environment may produce black holes of $30-90\msun$.

The theoretical studies of the evolution of very massive stars
have been rekindled   by the 
discovery of the ULXs, with the result that 
black holes of $\sim 100\msun$ are more common than
earlier works suggested.
Isolated solar--abundance stars may have masses up to $150\msun$,
but in dense environments they may coalesce to form stars as massive
as $1000\msun$ \citep{Yun08}. 
Recent studies on the mass loss from hot, massive stars
(\citealp{Pul08}, and references therein) have shown that
the occurrence of weak and/or clumpy winds may reduce 
the mass loss up by  a  factor $10$, 
thus allowing the star to retain a higher fraction of its 
original mass at the end of its life. 
Also  lower metallicities  entail weaker wind mass losses
\citep{Heg03}. Zero metallicity (Population~III) stars 
may be significantly more massive than solar--abundance stars
\citep{Ohk06}.
All these factors determine the mass of the final black hole,
that may range from $\sim 70\msun$ for a solar--abundance progenitor
\citep{Yun08}, up to $500\msun$ for low metallicity stars
\citep{Ohk06}.
Observations indicate that the low metallicity 
scenario is more likely for N10:
the Cartwheel galaxy is metal--poor, and although the  available measurements
\citep{Fos77} are not at the exact location of N10, they  should 
nevertheless be
representative of the average abundance in the ring. 
We therefore conclude that the interpretation of N10 as a black hole
of $\sim 100\msun$ suggested by the X--ray data is consistent
with the theoretical predictions of the evolution  of massive stars
in a  dense environment.

In summary, if ordinary stellar evolution is the correct scenario
to form the black hole, N10 
could be  an extreme  High Mass X--ray Binary (HMXB).
The mass accretion rate $\dot M$ on the black hole 
may be inferred from the luminosity $L$ via Equation~\eqref{e-eff}, 
and is of the order of ${\dot M}\simeq10^{-6}~\msun\yr^{-1}$.
This value is comparable with the loss rate of 
massive (i.e., few tens solar masses, \citealp{Fra02}) 
donor stars on a thermal time scale.
The accretion flow from the donor star to the BH is most likely to
occur through a  Roche lobe overflow; the BH capture of a wind 
blown by the companion is less favoured, since the small capture 
radius would require an unlikely  strong mass loss from the companion.

Can  the observed decay of the luminosity of the source be explained in the
framework of the accretion model? 
The longest characteristic time of an accretion disc is the so--called
``viscous time'' $t_{\rm visc}$, i.e.  the characteristic time taken
by the disc  to  adapt to new conditions. 
For  a standard $\alpha$-disc  
orbiting a black hole of $100\msun$ with an accretion rate
${\dot M}\simeq10^{-6}~\msun\yr^{-1}$,
$t_{\rm visc}\simeq 10^2-10^3\s$ 
\citep{Fra02},
much shorter than the time scale of the variability of
N10 (few years). Therefore, the observed  variability cannot be due to
a sort of disc instability, which would affect the disc and the 
luminosity of the source over a time scale $t_{\rm visc}$.
The simplest alternative  is that the variability is due to a change of
the mass transfer rate from the donor star. This may be due to
an intrinsic decay of the mass loss from the donor, but also  to 
other effects occurring at the inner Lagrangian point, since the instantaneous
mass flow here  is
very sensitive to the relative sizes of the donor star and the Roche lobe
\citep{Fra02}.
New observations of this interesting source would help to settle the question.

\bigskip

We also explored the possibility that N10 is a young supernova
strongly interacting with the circumstellar medium. Both the
interpretation of this model and the best--fitting
spectral parameters are consistent with this view.
A new flux brightening of the source in the future would rule out this 
hypothesis.


\section*{Acknowledgments}
We acknowledge financial support from INAF through grant PRIN-2007-26.
We thank Luca Zampieri for a careful reading of the manuscript and
for his useful observations. 
We also thank an anonymous referee for her/his comments that helped
to improve the paper.


\clearpage

\nocite{}
\bibliographystyle{mn2e}
\bibliography{refs}


\clearpage


\clearpage


\begin{table*}
\centering 
\begin{minipage}{140mm}
\caption{{\sl Chandra} observation log of the Cartwheel galaxy.}
\label{t-data}
\begin{tabular}{ccccc}
\hline
Observation & 
Exposure time & 
Count rate &
Mode & 
Start date \\ 
identification & 
(before) after cleaning & 
($0.3-7\keV$) & 
 & 
 \\ 
 & 
(ks) & 
($10^{-3}$~cts/s) &
 & 
(yyyy/mm/dd) \\
\hline
2019 & (76.5) 76.1 & $5.3$  & FAINT & 2001/05/26  \\
9531 & (52.8) 48.0 & $3.6$  & VFAINT & 2008/01/21  \\
9807 & (49.8) 49.5 & $1.2$  & VFAINT & 2008/09/09  \\
\hline
\end{tabular}
\end{minipage}
\end{table*}


\begin{table*}

\centering
\begin{minipage}{140mm}
\caption{
Summary of the absorbed power--law model ($\tt wabs*powerlaw$).
} 
\label{t-plaw}
\begin{tabular}{cccc}
\hline
Obs. id.   &  
Normalisation       &
$F_{0.5-10\keV}$    &
$L_{0.5-10 \keV}$    \\
                          &
$10^{-5}$ XSPEC units     & 
($10^{-14}\ergs\cm^{-2}$)  &
($10^{40} \ergs$)
\\
\hline
2019                       & 
$1.24_{-0.27}^{+0.36}$      &
$6.60$   &
$11.8$
  \\                 
9531                       & 
$0.91_{-0.21}^{+0.27}$      &
$4.81$  &
$8.6$
  \\
9807                       & 
$0.29_{-0.08}^{+0.11}$     &
$1.56$  &
$2.8$
\\
\hline
$n_H=3.67_{-0.93}^{+0.99}\times 10^{21}\cm^{-2}$ &
$\Gamma=1.88_{-0.24}^{+0.25}$ &
$C/{\rm dof} = 100.8/101$ &
 \\
\hline
\end{tabular}
\end{minipage}
\end{table*}

%

\input{table3.tex}



\begin{table*}

\centering
\begin{minipage}{140mm}
\caption{
Summary of the 
	absorbed APEC model {\tt wabs* apec}.
}		
\label{t-apec}
\begin{tabular}{cccc}
\hline
Obs. identification no.    & 
Normalisation              &
$F_{0.5-10\keV}$           &
$L_{0.5-10\keV}$          \\
						  & 
$10^{-5}$  XSPEC units     &
($10^{-14}\ergs\cm^{-2}$)  &
($10^{40}\ergs $)
\\
\hline
2019                       & 
$4.3_{-1.2}^{+0.9}$        & 
$5.55$         &
$9.9$           \\
9531                       & 
$3.1_{-0.9}^{+0.7}$     & 
$4.04$  &
$7.2$                   \\
9807                       & 
$1.0_{-0.2}^{+0.3}$     &
$1.31$  &
$2.3$                    \\
\hline
$n_H =2.8_{-0.6}^{+0.8}\times 10^{21}\cm^{-2} $ &
$k_B\,T = 5.1_{-1.6}^{+3.1}\keV $ &
$Z=0^{+0.2}$ &
$C/{\rm dof} = 95.9/100$ \\
\hline
\end{tabular}
\end{minipage}
\end{table*}


\clearpage

\bsp

\label{lastpage}

\end{document}

%% file: table3.tex



\setcounter{table}{2}
\begin{table*}

\begin{turn}{90}


\begin{minipage}{261mm}
\caption{
Summary of the accretion disc models:
MCD={\tt wabs*diskbb}, MCD+C={\tt wabs*(diskbb+comptt)}, 
Kerr={\tt wabs*kerrd}  Slim={\tt wabs*Kawaguchi}. 
The first group of rows shows the best--fitting parameters for three  observations
separately.
The second groups presents the values of the parameters common to all the observations, and the other parameters discussed in the text.
Fluxes and luminosities are calculated in the spectral band
$0.5-10\keV$. $K_D$ and $K_C$ stand for the disc and {\tt comptt} normalisation, respectively. 
$f$, $\eta$ and $\theta$ are the hardening parameter, the accretion efficiency and the 
inclination of the disc.
The luminosites are total, so the luminosity of the MCD+C model also includes the
contribution of the Compton corona. 
}
\label{t-acc}

\begin{tabular}{|c|ccc|ccc|ccc|ccc|} 
\hline
Obs. id. & \multicolumn{3}{|c|}{MCD}  & \multicolumn{3}{|c|}{MCD+C}  & \multicolumn{3}{|c|}{Kerr}  & \multicolumn{3}{|c|}{Slim}   \\\hline
         & $k_BT_{\rm in}$ &  Flux & L$_X$ & $k_BT_{\rm in}$ & Flux & L$_X$ & $\dot{M}$ & Flux & L$_X$ & $\dot{M}$ & Flux & L$_X$  \\
         &  ($\keV$) & $10^{-14}$ &  $10^{40}$ &  ($\keV$) & $10^{-14}$ & $10^{40}$ & $10^{20}~\rm g\s^{-1}$ & $10^{-14}$  & $10^{40}$ & $L_{\rm Edd}/c^2$  &  $10^{-14}$ & $10^{40}$  \\
         &   & $\ergs\cm^{-2}$ &  $\ergs$ &   & $\ergs\cm^{-2}$ & $\ergs$ &  & $\ergs\cm^{-2}$  & $\ergs$ &  &  $\ergs\cm^{-2}$ & $\ergs$  \\
\hline 
2019 &$1.33_{-0.17}^{+0.22}$  & $4.63$ & $4.1$ & $0.53_{-0.12}^{+0.11}$ & $5.32$ &
$8.6$ &  $7.1_{-0.7}^{+0.8}$ & $4.84$ & $4.3$ & $39.4_{-23}$         &  $11.5$ & $5.8$  \\
9531 &$1.21_{-0.16}^{+0.17}$  & $3.09$ & $2.7$ &  $0.36_{-0.09}^{+0.16}$ & $3.63$ &
$6.3$ &  $4.8_{-0.7}^{+0.8}$ & $3.28$ & $2.9$ & $21.4_{-12}^{+523}$  &  $8.0 $  & $4.0$ \\
9807 &$0.89_{-0.11}^{+0.14}$  & $0.88$ & $0.8$ & $0.09_{-0.04}^{+0.05}$ & $1.13$ & 
$2.0$ & $1.5_{-0.3}^{+0.4}$ & $0.94$ & $0.85$ & $5.1_{-1.2}^{+79}$   &  $2.3$   & $1.1$ \\
\hline
$n_H (10^{21}\cm^{-2})$    &\multicolumn{3}{|c|}{$1.80_{-0.6}^{+0.6}$} & \multicolumn{3}{|c|}{$1.7_{-0.7}^{+0.8}$} & \multicolumn{3}{|c|}{$2.26_{-0.60}^{+0.64}$ } & \multicolumn{3}{|c|}{$3.6_{-0.9}^{+0.7}$} \\
$K_D$       & \multicolumn{3}{|c|}{$7.4_{-3.3}^{+5.2}\times 10^{-4}$ } & \multicolumn{3}{|c|}{$7.6_{-3.9}^{+10.6}\times 10^{-3}$ }  & \multicolumn{3}{|c|}{ $1$ (fixed) } &   \multicolumn{3}{|c|}{ $6.72\times 10^{-9}$ (fixed)} \\
$K_C$       & \multicolumn{3}{|c|}{ - }                   & \multicolumn{3}{|c|}{$2.27_{-0.6}^{+0.6}\times 10^{-7}$ }      & \multicolumn{3}{|c|}{ - } &   \multicolumn{3}{|c|}{- } \\
$T_C$       & \multicolumn{3}{|c|}{ - }                   & \multicolumn{3}{|c|}{$50\keV$ (fixed)}      & \multicolumn{3}{|c|}{ - } &   \multicolumn{3}{|c|}{- } \\
$\tau$      & \multicolumn{3}{|c|}{ - }                   & \multicolumn{3}{|c|}{$0.5_{-0.2}^{+0.4}$}   & \multicolumn{3}{|c|}{ - } &   \multicolumn{3}{|c|}{ -} \\
$\theta$  & \multicolumn{3}{|c|}{$0$}        & \multicolumn{3}{|c|}{$0$}       & \multicolumn{3}{|c|}{$0$}      & \multicolumn{3}{|c|}{$0$} \\
$\alpha$    & \multicolumn{3}{|c|}{ - }                   & \multicolumn{3}{|c|}{ - }                      & \multicolumn{3}{|c|}{ - } &   \multicolumn{3}{|c|}{$0.56_{-0.37}$} \\
$M_\bullet$ & \multicolumn{3}{|c|}{ - }                   & \multicolumn{3}{|c|}{ - }                      & \multicolumn{3}{|c|}{$92.8_{-27.7}^{+32.5}$} &   \multicolumn{3}{|c|}{$495_{-340}$} \\       
$ f $     & \multicolumn{3}{|c|}{$2.0$}      & \multicolumn{3}{|c|}{$1.7$}     & \multicolumn{3}{|c|}{$1.7$}    & \multicolumn{3}{|c|}{$1.7$} \\  
$\eta$    & \multicolumn{3}{|c|}{$0.06$}     & \multicolumn{3}{|c|}{$0.06$}    & \multicolumn{3}{|c|}{$0.4$}    & \multicolumn{3}{|c|}{$0.06$}  \\
\hline
$ C/{\rm dof} $     & \multicolumn{3}{|c|}{$92.8/101$}      & \multicolumn{3}{|c|}{$88.5/99$}     & \multicolumn{3}{|c|}{$93.2/101$}    & \multicolumn{3}{|c|}{$96.7/100$} \\
\hline
\end{tabular} 
\end{minipage}


\end{turn}

\end{table*}
